\documentstyle[12pt,aaspp4]{article}

\begin{document}

   \title{The ESO K$'$-Band Galaxy Survey \\
	  Galaxy Counts\footnote{Accepted for publication in  The Astronomical Journal}}

   \author{P. Saracco$^1$, A. Iovino$^2$, B. Garilli$^1$, 
D. Maccagni$^1$ and G. Chincarini$^{2,3}$}
\affil{$^1$ Istituto di Fisica Cosmica del CNR, Via Bassini 15, 20133 Milano, Italy}
\affil{$^2$ Osservatorio Astronomico di Brera, Via Brera 28, 20121 Milano,
Italy}
\affil{$^3$ Dipartimento di Fisica, Universit\'a di Milano, Via Celoria
16, 20133 Milano, Italy}
% \authoraddr{address}

\vskip 2truecm
\slugcomment{$^*$Based on observations collected at the European Southern
Observatory, La Silla, Chile.}
\vskip 2truecm

\received{26 Sept. 1996}
%\journalname{Astronomical Journal}

   \begin{abstract}
We present $K'$-band galaxy counts  in the magnitude
range $13<K'<20$,
obtained from two independent surveys
near the South Galactic Pole, covering an  area of $\sim40$ arcmin$^2$ 
and  $\sim170$ arcmin$^2$, including
 $\sim200$ and $\sim1000$ objects respectively up to a magnitude limit
$K'\sim 20.0$ and $K'\sim19.0$.
At magnitudes $K>18.5$ we surveyed an area more than 7 
times wider that covered by previous $K$-band surveys.
Our counts are quite in good agreement with those in the literature
 at magnitudes brighter than $K\sim17$, while are systematically 
lower  at fainter magnitudes.
We confirm a change in the slope of the $dlogN/dm$ relation 
at $K'\sim17$ from 0.57  to 0.35,
but do not find the bump shown by other $K$-band surveys in the
magnitude range $16<K<20$.
Furthermore 10\% of the galaxies brighter than $K'\sim18$ have $B-K'<3$.
We suggest that  these blue objects represent a population of sub-$L^*$ nearby 
evolving galaxies.

   \end{abstract}

\section{Introduction}
Present day ground based  instrumentation allows to count galaxies
 fainter  than B=27 (Tyson 1988; Metcalfe et al. 1995) and the 
resulting $dlogN/dm$ relation shows an  excess of galaxies with 
respect to the predictions of a no-evolution model. 
Several authors have tried to explain 
such an excess suggesting some mild luminosity 
evolution, or more complex mechanisms
(White 1989; Broadhurst et al. 1992; Cowie et al. 1991).

Recent results from HST observations  cast new light
on this problem. Driver et al. (1995), Glazebrook et al. (1995)
and Abraham et al. (1996) show that while 
elliptical and early-type spiral number counts are in good agreement
with the predictions of no-evolution models,
late-type/irregular galaxies show an excess with respect to such 
predictions. Driver et al. (1996) also find that most of the faint
blue galaxy excess at $B\sim23.5$ is due to a population
of evolving dwarf  galaxies at $0.2<z<0.5$, even if it is 
unclear  which kind of evolution  they should undergo. 
In general, galaxy counts in the visible bands suggest that late 
Hubble types undergo some evolution  at redshifts  $z\le 0.5$. 
It should be noted, however, that optically selected samples 
tend to favor  star forming galaxies (i.e. 
evolving galaxies) and this is even more true in the B band.

Near-infrared (IR) selection may provide significant advantages over optical selection in the 
study of galaxy evolution:  k-corrections are small and nearly
 independent on  galaxy type (Cowie et al. 1994);
the sample is less biased towards star-forming galaxies; 
 the comparison models are simpler. 
Already from the 
first survey (Cowie et al. 1990), it was immediately realized that 
 K--band counts do not require  galaxy
evolution. 
Such a discrepancy between $B$--band and $K$--band counts has been 
 confirmed  by all the subsequent K--band surveys, which have extended 
the counts down to $K\sim22.5$  (Gardner et al. 1993; Cowie et al. 1994; 
Soifer et al. 1994; McLeod et al.  1995) and to $K\sim24$  
(Djorgovski et al. 1995). 
The signatures of evolutionary changes are however present also in the IR counts:
 K-band counts flatten at  magnitudes $K\sim17$ (Gardner et al. 1993) 
and a blueing of the galaxies occurs at fainter magnitudes
(Gardner et al. 1993; Cowie et al. 1995). 
These two observational evidences imply that field galaxies 
undergo substantial changes at $K\sim17$-18,
which is the  expected magnitude of an  $L^*$ galaxy at $z\sim0.3$.
Colless et al. (1993; 1994) and Griffiths et al. (1994a,b) have shown
that, at these redshifts, there is an
increase of  starburst activity, mainly associated with tidal
and merging phenomena, while Driver et al. (1996) show  that a large fraction of dwarfs
present at $z\sim0.5$ has subsequently  faded to obscurity.

Thus, our understanding of galaxy evolution would greatly benefit from
a {\it large, IR-selected} sample of galaxies in the magnitude 
range $15<K<20$.
Such a sample is still missing.
 The deep IR surveys 
go fainter than $K\sim18$ (Cowie et al. 1990; Cowie et al. 1994;
Soifer et al. 1994; McLeod et al. 1995;  Djorgovski et al. 1995;
Moustakas et al. 1997) but
cover  small areas (from $\sim1$ to $\sim20$ square arcminutes), 
thus giving  small numbers of galaxies
at magnitudes $K<17$-18.
On the other hand, the  limiting magnitudes of shallower surveys 
(covering larger
areas) are either brighter than $K=16$ 
(Gardner et al. 1996; Huang et al. 1997) or fall  in the range
$17<K<18$ (Glazebrook et al. 1994; Gardner 1995a), which implies
a possible  incompleteness at these magnitudes.

At the end of 1993 we started a project aimed at
selecting an IR  galaxy sample deep and large
enough to  study  galaxy evolution.
The project consists of two surveys:
the  ESOKS1 (ESO $K'$-band Survey 1) limited to $K'=20$ over 
$\sim40$ arcmin$^2$ of sky and mainly designed  to define  faint galaxy
counts and to study faint objects,  and the ESOKS2 limited to 
$K'\sim19$ over a  region of $\sim170$ arcmin$^2$, making up the
bulk of our sample for $K'\leq19$.

In this paper we describe the two surveys and  the  
analysis of the ensuing galaxy counts.

%
%__________________________________________________________________
\section{The Surveys}
\subsection{Field Selection}
The  fields observed in the two independent surveys were selected
in two slightly different ways and cover unconnected areas.

\paragraph{ESOKS1:}
We  selected 12 fields according to the following
criteria:
\begin{enumerate}
\item each field had to be at high galactic latitude ($b<-45^o$),
in order to minimize star contamination and galactic extinction;
\item all fields had to include two medium brightness stars
($17.5\le b_j\le19$), in order to optimize the stacking of the different
IR frames;
\item the fields had to be free from clearly visible objects in 
the ESO--SRC sky   survey plates (apart from the two stars of point 2).
\end{enumerate}
This last requirement was intended to avoid having a significant part
of the field covered by bright galaxies, being the ESOKS1  aimed
 at defining galaxy counts at magnitudes fainter than  
$K'=16$  (down to $K'\sim20$) and to construct a faint object catalog.
Obviously, this criterion introduces a bias against $b_j<20.5$
objects.
From the B-band galaxy counts (e.g. Tyson 1988; Maddox et al. 1990; 
Jones et al. 1991; Metcalfe et al. 1991; Bertin \& Dennefeld 1997)
we would expect to find  7-9 such galaxies in the 
$\sim$40 arcmin$^2$ of the survey.
According to the $B-K'$ color distribution  of the $B$-selected galaxy 
sample  of Bershady et al. (1994), only 1 of these  
galaxies should be in the range $16<K'<17$ and only 0.2 galaxies
in the range $17<K'<18$.
Since the expected number of galaxies fainter than $K'\sim16$  is $\gg 1$
in the ESOKS1 area, this bias is negligible.

\paragraph{ESOKS2:}
The purpose of this second survey was to explore a larger area,
 reaching magnitude limits brighter than  ESOKS1
(down to $K'\sim19$).
We randomly selected 51 fields according to the first two selection
criteria and dropped the last criterion in order to
construct  a  sample  unbiased with 
respect to bright and blue objects.
However, because of observing time scheduling, it was not 
possible to select all the fields at high galactic latitude.
In particular 13 fields are at $-45^o<b<-20^o$,
 while the remaining 38 fields are at high galactic latitude ($b<-45^o$).
  
In Table 1 and 2 we give
the right ascension, declination and the galactic latitude
of the center of each field, and the 
integration time ($t_{exp}$) for each of the ESOKS1 and ESOKS2 fields 
respectively.

%-------------------------------------------------------------
\subsection{Observations}
Observations were made at the  ESO 2.2m  telescope equipped
with the IRAC2B infrared camera (Morwood et al. 1992).
This camera is based on a NICMOS3 array (Hodapp et al. 1992)
of 256$\times$256 HgCdTe diodes.
With objective C, the pixel size is 
0.5 arcsec, thus providing a field of view of
$2.1\times2.1$ arcmin and 
a reasonable sampling of the object profiles under average seeing conditions.

The data were obtained using the $K'$ filter ($\bar\lambda=2.15 \mu$m and 
$\Delta\lambda\sim0.32$).
Tests performed at the ESO 2.2 m telescope show that through the $K'$ 
filter the sky+telescope
background   is 35\% lower than through
the standard $K$ filter, yielding a 20\% reduction in  sky noise
(Lidman et al. 1995).

%\subsubsection{Galaxy Fields}
The 12 ESOKS1 fields  were observed on October 5 through 7, 1993.
The seeing was in the range 1-1.5 arcsec.
One field was observed for  $t_{exp}=36$ min, 
7 fields for $t_{exp}=54$ min, 2 fields for $t_{exp}=72$ min and
the remaining 2 fields for $t_{exp}=162$ min.
The 51 fields of the ESOKS2 were observed during a run of 5 
nights  starting on August 25, 1995.  
The seeing was in the range  0.9-1.3 arcsec. 
The exposure time was 36 minutes for 50 out of the 51 observed fields,
and  18 minutes for the remaining one.

Due to the brightness of the background  ($\sim$12.6 $K'$
mag arcsec$^{-2}$ from a combination of sky, telescope and instrument,
Morwood et al. 1992), extremely accurate flat-fielding is required
(about 1 part in 10$^4$ in each pixel) to reach a magnitude limit of
$K'\sim20$.
In order to achieve this result the "shift and stare" observational
technique (Tyson 1989) was adopted. 
 We obtained 9 frames around the center of each
field by moving the telescope by 10 arcseconds each time in a well
defined grid pattern.
In order to avoid pixel saturation effects, each frame is the average
of 12 exposures of 10 s each, and the central $1.8\times1.8$ arcmin
of each image, obtained from the stacking of the 9 mosaic frames, 
 has an exposure time of 18 min.
 Longer exposure times
 are obtained by repeating the pattern.
This technique allows to have sky 
limited exposures while keeping the overheads to a reasonable 
small fraction of the total observing time.

14 fields at $b>-60^o$ out of the 51 fields of the ESOKS2 have
also been observed  in  the $B$ and $R$ bands (they are marked with
* symbol in Table 2).
These observations were obtained at the ESO 0.9 m Dutch telescope
in August 1995, with exposure times of 60 minutes in the $B$-band
and 30 minutes in the $R$-band.
The pixel size is 0.44 arcsec and the  seeing, throughout the 
observations, was in the range 1.3-1.8 arcsecs.
Here we  present the colors of the galaxies in these fields in the framework of the $K$ galaxy 
counts.

\subsubsection{Photometric Accuracy}
Absolute photometry was obtained by observing standard stars taken
from the lists of Elias et al. (1982) and Legget \& Hawkins (1988) with
magnitudes in the range $8<K<11.5$.
During each night we observed from six to eight standard stars evenly
spaced in time. 
Each individual frame was the average of 12 exposures of 2 seconds
each.
We derived K' magnitudes from the relation
\begin{equation}
K'-K=(0.20\pm0.04)(H-K)
\end{equation}
(Wainscoat \& Cowie 1991).
The typical scatter in the zero-point  is 
0.05 mag for the  ESOKS1, and 
on the order of 0.03  mag for the ESOKS2.

\section{Data Reduction}
Data reduction and analysis was  based on the IRAF 
data reduction package.
After correction for the dark current, for each frame we constructed its own median sky-flat,
using an adequate number of  adjacent frames. 
Since the  sky color changes over time scales on the order of
$\sim15$ min (e.g. Glazebrook et al. 1994), 
no more than 7-8 time-adjacent frames can be used to
construct the flat-field of each frame.
On the other hand,  a good  sampling of the pixel response
is only obtained  from a sufficiently high number of frames.
We found that the best compromise is to combine
 6-7 time-adjacent frames to
obtain the sky flat of each image, taking 
care to remove objects and cosmic rays using a 
median sigma--clipping rejection algorithm. 
Each frame was then flat-field corrected
and a constant sky value subtracted. The resulting sky-subtracted flat-fielded frames
were registered and combined with median weights.
In order to optimize the co-adding, each frame was re-sampled
with smaller pixel size taking care to preserve the flux.
In the resulting final frames,  both the  pixel-to-pixel variations and
the  large scale  spatial variations   are on the
order of 10$^{-4}$ counts per pixel.

%   \begin{enumerate}
%      \item The conditions for the stability of static, radiative
%   \end{enumerate}

\section{Detection, Completeness and Photometry: Simulations}

Source detection was performed using FOCAS (Jarvis \& Tyson 1981; Valdes 1982).
As the detection algorithm is based on
two main
parameters, detection threshold and minimum number of connected 
pixels  satisfying the threshold criterion and defining the source,
detectability primarily depends on the S/N ratio of the pixels defining
 an object and on their being contiguous.
Optimization of the two algorithm parameters allows to push the 
detection to 
the faintest limits, while keeping low the number of spurious detections. 
Incompleteness increases as low surface brightness objects fall below 
the detection threshold.
In order to evaluate the corrections for incompleteness and the
contamination due to spurious detections, we carried out a set of simulations.

We  constructed ten
background frames (noise frames) by combining the appropriate number of 
different flat-fielded 2 min frames  for each exposure time.
Cosmic rays and objects from the different frames were removed by taking the median of the pixel 
intensity in each frame. 

Simulated galaxies were then  added to each background frame. 
We  divided the magnitude range $18.0<K'<20.0$
into four 0.5 mag wide bins, and  magnitudes were randomly assigned to
 each galaxy within each bin  so that a log(number)-magnitude relation 
with slope 0.6 was globally  reproduced.
For each bin, we simulated 300 galaxies, assuming 
a morphological mix of 35\% early type galaxies (E+S0) and
65\% late type galaxies (Guiderdoni 1991).
Early type galaxies were assumed to have a 
de Vaucouleurs profile and spirals a pure exponential one;
ellipticities and position angles were randomly assigned to each
galaxy and the effect of clustering was not included.
The apparent size of each galaxy was rescaled following the
diameter-distance relation, assuming that all galaxies have 
 luminosity $L^*$.
Each galaxy was convolved with a Moffat
function with a FWHM of 1.25 arcsec, which is a fairly
good approximation to our seeing profile. 
A galaxy having an absolute
magnitude on the order of $L^*$ and an apparent magnitude of
$K'\sim18.5$, has an apparent effective radius $\theta<0.5$ 
arcsec, irrespective of its intrinsic profile. 
Therefore, at magnitudes fainter than $K'\sim18.5$ most of the objects
are pointlike in our frames, and both detection and completeness
do not depend on galaxy profiles and/or 
morphological mixture.
Then, we added each set of 300 galaxies
to the 10 noise frames (30 galaxies per frame per magnitude bin).
This procedure was repeated for the different exposure times.

To increase the signal-to-noise ratio, and therefore the detectability
of the sources, we convolved the simulated images with a pass-band filter (boxcar) having a running window slightly
greater than the seeing disk (see  Bernstein et al. 1995; 
Banchet et al. 1997; Mamon 1995).
We  applied the FOCAS detection algorithm to the convolved images
trying different values of detection threshold and area: the
best results (i.e. the lowest number of spurious detections
and a high completeness) were obtained
with a detection threshold of 3$\sigma$ per pixel and a minimum
detection area $\sim50\%$ greater than the seeing disk.
In Table 3 the derived completeness correction factor $c$ 
is shown for each 
magnitude bin as a function of the exposure time. 

Three different types of magnitudes can be computed with FOCAS: 
isophotal ($K'_{iso}$), fixed aperture ($K'_{apr}$) and 
total magnitude ($K'_{tot}$).
The isophotal magnitude is calculated within the 
detection isophote, while
the total magnitude  is derived by growing the object several
pixels in all directions around the original detection isophote
and measuring the flux above the sky inside this expanded region.
We chose apertures of 3 and 5 
arcsec in diameter, i.e. at least twice the largest FWHM of the
seeing disk in our images. 
 To decide which of these
different quantities  best approximates  the 
"real" magnitude of the galaxies, we have compared them with 
the  magnitudes $m_{true}$ of the simulated galaxies.
Even if the best agreement is obtained between $K'_{tot}$ and
$m_{true}$,  FOCAS total magnitudes tend to 
systematically underestimate the flux of the sources fainter than
$K'\sim17.5$ (ESOKS2) and $K'\sim18.5$ (ESOKS1).
By comparing the estimated and the simulated
flux we computed the correction  to be applied to $K'_{tot}$: 
the applied corrections are 0.06, 0.15, 0.25 and 0.35 mag for sources
in the magnitude range $17.5<K'<18$, $18<K'<18.5$, $18.5<K'<19$ 
and $K'>19$ respectively.
After this correction the  scatter of
$\Delta m=K'_{tot}-m_{true}$  is $\sigma_{\Delta m}\sim0.2$
for galaxies with 19.0$<K'_{tot}<$20 in the case of $t_{exp}\ge54$ min
(ESOKS1).
The same scatter  is present for sources in the range $18<K'_{tot}<19$
for $t_{exp}\le36$ min (ESOKS2).
The scatter decreases rapidly going to brighter magnitudes and it is
 $\sim 0.1$ mag at $K'<19$ ($t_{exp}\ge54$ min) and 
$K'<18$ ($t_{exp}=36$ min).
Therefore we consider 0.2 mag as our
typical uncertainty on the faintest magnitudes.

\section{Image Analysis}
\subsection{Source Detection and Photometry}
The detection procedure described above was applied to the real 
frames.
Over the $\sim$40 arcmin$^2$ of the ESOKS1, 198 objects with magnitudes 
$K'_{tot}\le20$ were detected, while 998 sources brighter
than $K'_{tot}=19.5$ were detected over the $\sim$170 arcmin$^2$
of the ESOKS2.
In Fig. 1 and Fig. 2 the $K'_{tot}$ magnitude distributions for the two 
samples  are shown. In Fig. 2, also the $K'_{3''}$ and $K'_{5''}$ magnitude distributions
are plotted.
The distributions of the three different magnitude estimates are
in good agreement down to $K'=18$.
At fainter magnitudes the three distributions are slightly different
since both the 3 and 5 arcsec apertures tend to
underestimate the source  flux, even if the differences are within 
1$\sigma$.
Thus we do not expect a significant dependence of the number counts
on the particular method chosen to estimate the magnitude.
From now onwards we will use  $K'_{tot}$ magnitudes (corrected as
specified in Section 4)
unless  explicitly stated.

In the 14 fields observed also in the $B$ and $R$ bands, there
are 398 $K'$-detected objects: 312 of these objects are also
detected in both $B$ and $R$.

\subsection{Star-Galaxy Separation}
There are several methods  to
separate stars from extended objects, including the purely
statistical one, based solely on star counts models.

Gardner (1995b) and McLeod et al. (1995) base their classification
on source colors: given 3 filters, $B$, $K$ and $X$, where $X$ is
a third red band, in the color--color diagram $B-X$ vs $X-K$ main
sequence stars define a specific locus, while normal galaxies
are expected to lie in the lower part of the diagram.
On the contrary, Metcalfe et al. (1991) base their classification
on the isophotal area: at a given magnitude, stars should have
a smaller isophotal area than galaxies.
Having observed the 14 fields at low galactic latitude also in the 
optical bands with a much higher S/N, we checked the applicability 
of both these methods to our data.
We  ran the FOCAS classifier (which takes into account the 
first moment of the radial distribution of the light
of the source) on both the $B$ and $R$ images
and considered as classified those sources for which
the classification coincided.
The reliability of this classification was checked on the
brighter sources by visual inspection. We have coincident classifications
for 85\% of the sources down to $K'=18$ (214 sources).
In Fig. 3 we present the $B-R$ vs $R-K'$ color--color plot
for the  classified objects (stars represent stars and 
open circles galaxies), and for the unclassified objects 
(crosses).
Colors were obtained using $B$, $R$ and $K'$ 5 arcsec aperture
magnitudes: aperture magnitudes avoid problems introduced by the 
different sampling of the 
galaxy profile in each band due to the different depth of the images.
A 5 arcsec diameter aperture gives reliable colors being at least
3 times larger than the typical FWHM both in the IR and optical
images.
It is clearly visible that in the ($B-R$)-($R-K'$) plane 
 stars and galaxies are not distinguishable, 
the two categories being heavily intermixed.
Thus, in our case, a classification based on this method, which
would be applicable only to 14 out of the 51 fields,  would imply
the rejection of a non-negligible fraction of galaxies (mainly
blue galaxies) and a strong star contamination.
However, we note that the use of the $I$,  as in the case of Gardner et al. (1995b), instead of the $R$ filter reduces such 
effect, even if  cannot completely eliminate it  (see Fig. 8 in 
Cowie et al. 1994).

The method of the $K'$ isophotal area would have
 the advantage
of being applicable to all our fields.
In Fig. 4 $K'$ isophotal area vs $K'$ magnitude is plotted.
The starred symbols are the objects independently classified as 
stars while circles are galaxies.
Stars separate very well from galaxies down to $K'\sim15$, but fainter
than that, galaxies  tend to mix up with point-like sources.
Given the resolution of our IR images, this method would tend to
discard the more compact galaxies.

We also compared the FOCAS classification obtained from the optical
images with the one from the IR frames.
In Fig. 5 the ratio between the number of objects classified as stars on 
the IR frames ($S_K$) and that on the $B$ and $R$ frames ($S_{BR}$) as a 
function of $K'$ magnitude is shown: 
the IR based  classification systematically underestimates  
 the number of stars.
As a result of this analysis, we  decided to use two different 
and independent methods to correct the number counts for star contamination.
The first method is based on the FOCAS classification performed on the
IR images over the whole range of magnitudes and corrected in each magnitude bin for the 
underestimate shown in Fig. 5. 
At magnitudes fainter than $K'=18$ we assumed the same correction 
 obtained in the bin $17<K'<18$. 
The second  method is based on the galaxy model 
 of Cohen (1993) and has been used only for the fields of ESOKS2.
In this case, since star counts are strictly dependent on galactic coordinates, we grouped our fields in three galactic latitude bins: 
low ($b>-40^o$), high ($-60^o<b<-40^o$)  and pole ($b<-60^o$), 
and for each of them we estimated the expected number of stars down
to $K'\sim20$.
It is worth noticing that the predicted star counts  agree very well 
with the number of observed stars in the optical images.
A more detailed discussion of the stellar contamination results obtained
using this galaxy model is presented  in Cohen and Saracco (1997).

\section{Results}
\subsection{Galaxy Counts}
The second selection criterion for our fields required the presence 
of 2 medium 
brightness stars to optimize the stacking of the frames.
Although these objects were classified as stars in the ESO-SRC based
 COSMOS
catalog, a fraction  ($\sim30\%$) turned out to be galaxies, 
both on the basis of the FOCAS
 classification and on a visual inspection of the frames.
In order to prevent a bias toward bright galaxies, such galaxies have 
not been taken into account in  deriving  the galaxy number counts.

A small fraction of detections, 5\% in the ESOKS1 (9 objects)
 and 7\% in the ESOKS2 (71 objects),
 are sources lying  near the edges of a frame, for which
we cannot correctly  estimate  their magnitudes.
We can either reject such detections
reducing the useful area of the images (as in Driver et al. 1995),
or assume for them a given magnitude and  
star fraction distribution.
As in both cases some assumptions have to be made
(the area to be discarded in the first case, and the source flux
distribution in the second case), we have checked how much the
overall results are affected by adopting
one or the other of these approaches. 
We compared the results obtained adopting two different rejection areas
(a strip 4 times and 2 times the minimum detection radius all
around the images) or keeping all detections and assuming
that their
magnitude distribution and star fraction distribution 
reflect that obtained from all the other sources.
The maximum discrepancy in the number density of sources
was less than 4\%.
Thus, we have retained all detections 
in order to maximize the number of 
objects and the total surveyed area.

Then we  corrected the object counts for spurious detections
and incompleteness, using the results of the simulations.
Finally, we  subtracted the contribution from the
stars according to the procedure described in \S 5.2.

The resulting galaxy counts for the ESOKS1 and ESOKS2 
are  listed in Table 4 and Table 5 respectively, where we report for each
magnitude bin the effective area, the raw counts including stars (Raw),
the counts corrected for  edge sources, incompleteness and star 
contamination according to the properly scaled FOCAS classification ($n$), the  counts $N$ per square degree and the 1$\sigma$
Poissonian error calculated as the square root of the raw counts.
For the ESOKS2 (Table 5), we also report the counts obtained by subtracting the number of stars given by Cohen's galaxy model ($n_{mod}$ and $N_{mod}$),
the raw number of sources (galaxies and stars, Raw$_{b<-75^o}$) and
the number
per magnitude bin per square degree ($C_{b<-75^o}$) detected in the 16 fields at
galactic latitude $b<-75^o$.
Since  we expect that the contribution of stars at these latitudes 
is negligible at magnitudes fainter than $K'=18$, these counts
represent an upper limit to our galaxy counts and at the same time
a good landmark for our star counts. 

In Fig. 6 the galaxy counts derived from the ESOKS1 and ESOKS2 are shown.
The agreement between our two independent surveys is very good, the
larger difference being within 1$\sigma$.
The counts  follow  a $dlogN/dm$ relation with a slope of  $\sim0.35$
in the magnitude range $17<K'<20$, both in the ESOKS1 and in 
the ESOKS2.
At brighter magnitudes ($K'<17$) the counts slope is $\sim0.57$, 
as derived from the ESOKS2 data.

\subsection{Field-to-Field Variations}
Galaxy counts differ from field to field.
The scatter on the number of galaxies detected in each field 
is $\sigma_{N_1}=4.7$ for ESOKS1 (here we
considered only the 7 fields with $t_{exp}=54$ min, in order
to have the same completeness limits) and $\sigma_{N_2}=4.9$ 
for ESOKS2 (in this case we excluded low galactic latitude
fields, because of the much higher star contamination).
Given the mean number of galaxies per field ($\bar N_1\sim13$ and
$\bar N_2\sim10$ for ESOKS1 and ESOKS2 respectively), we expect
a pure poissonian noise of $\sigma_{p_1}=3.6$ and $\sigma_{p_2}=3.1$ 
galaxies per field.
Another contribution to field to field variation is due to galaxy
clustering.
For an angular correlation function with a power--law form
$w(\theta)=A_w\theta^{-\gamma}$ and a circular window function of radius
$\theta_0$, the expected fluctuations due to clustering are:
\begin{equation}
\sigma_\omega=f(\gamma)w(\theta)^{1/2}\bar N 
\end{equation}
where $f(\gamma)\sim1$.
Brainerd et al. (1994) show that $log A_w\sim-0.3r_{lim}$
where $r_{lim}$ is the magnitude limit of the sample.
Assuming  $<r-K>\sim3$ mag we rescaled the values of $A_w$ estimated
by Brainerd et al. and extrapolated their results to $r_{lim}\sim22.5$.
We found $A_w\sim1.47$ (with $\theta$ in arcsec).
Since the area of each field ($\sim3.2$ arcmin$^2$)
corresponds to a circle of radius
$\theta_0\sim60''$  the expected field--to--field variation due
to clustering is $\sigma_w\sim3.1$ galaxies.
Combining $\sigma_\omega$ with $\sigma_p$ we expect 
$\sigma_{exp_1}=4.8$ and $\sigma_{exp_2}=4.4$ 
galaxies per field, both consistent with the measured values 
of $\sigma_N$.

\subsection{Galaxy Colors}
Visible-NIR colors and classifications of objects are available for the low
galactic latitude fields. 
At $K'\sim18$, the classification is reliable for 85\% of the IR sources:
204 out of the 214 sources brighter than $K'=18$ have been detected in B;
115 of them are stars, 66 are galaxies and the remaining 23 are 
unclassified sources.
In Fig.7 we show the $B-K'$ vs $K'$ diagram for the 66 galaxies
(open circles) the unclassified 
$K'\le18$ sources (crosses) and the remaining $K'>18$ objects
 (filled circles).

From Fig. 7 it is apparent that there is  a small but non-negligible 
fraction of  galaxies (9 galaxies) with   $B-K'\le3$ at $K'\le18$.
It is worth noticing that the IR photometric errors at these magnitudes
are not larger than $0.1$ mag and therefore the uncertainties 
in color estimation are also $\sim0.1$ mag.
These  galaxies  represent $\sim$10\% of our $K'\le18$ sample,
and their very blue colors suggest the presence of star-formation
activity.
The three bluest galaxies, with a
$B-K'$ colors of 1.8, 1.8 and 2.2 respectively,  also show very blue B-R 
colors, 0.9, 0.7 and 0.8 respectively.
The $B-K$ vs $M_K$ relation (Saracco et al. 1996) 
suggests that  galaxies brighter than $K'\sim18$ and
bluer than $B-K\sim2.5$ should be at a redshift $z<0.27$,
whilst the expected median redshift of a sample
limited to $K\sim18$ is $z>0.35$.
In other words, this population of blue  galaxies should be
intrinsically fainter than  $L^*$.
We suggest that they could  represent a  population of blue 
sub-$L^*$ galaxies  similar to those evidenced
 by Lilly et al. (1995) at $z<0.2$.
The absence of this population from  most of the other $K$-band
selected samples 
 is a natural consequence of the  blank field selection
criterion (e.g. Hawaii Deep Survey (HDS); ESOKS1; McLeod et al. 1995).
Moreover a star-galaxy separation based on the color--color diagram
may also introduce a bias against such blue galaxies.

In Fig. 8  the $B-K'$ color distribution for the $K'\le18$ ESOKS2
sample of galaxies (solid line) is compared 
to the HMDS sample (Gardner  et al. 1995).
10 ESOKS2 objects brighter than $K'=18$ have not been detected in the $B$ images.
Our $B$ limiting magnitude is $B\sim24$ and these objects are 
 fainter than $K'\sim17.5$. Thus
the 10 undetected sources are redder than  $B-K$=6
and our color distribution is complete for $B-K'\le6$ (see
also Fig. 7).
These 10 sources are represented in Fig. 8 by the shaded boxes
which represent the distribution of their $B-K'$ lower limits.
The HMDS was derived by surveying the SSA blank selected fields 
and the  Durham (Dur) fields, which are not  'blank'.
Since also the two HMDS subsamples include  lower limits  to
the $B-K$ color of some sources, we make use of a survival
analysis technique (Avni et al. 1980; Isobe et al. 1986; Feigelson \& 
Nelson 1985) to compare the colors of the ESOKS2 and the HMDS subsamples.
The color distributions of the two  HMDS subsamples shown in Fig. 8
are different at a very high confidence level: the probability
that the two samples have been  extracted  from the same parent 
population is less than $10^{-4}$ as given by the generalized 
Wilcoxon test.
We reach the same statistical conclusion by comparing the ESOKS2
sample with  the HMDS-SSA sample, a result which could be not surprising
given the  different  criteria used to select the HMDS-SSA fields
and the ESOKS2 fields.
On the contrary, the color distribution of the ESOKS2 sample is
in good agreement with the HMDS-Dur sample as confirmed by the
 probability of 15\%  that the ESOKS2 and the 
HMDS-Dur samples are drawn from the same population.

\section{Discussion}
The sample we selected is fundamentally different from previous
samples since it is composed by a high number of randomly selected 
and unconnected small fields, rather than by adjacent targets 
or unconnected large areas.
We compare our results with each one of the 
samples present in the literature.
We show that our counts are systematically lower
 at magnitudes fainter than $K'=17$  even if the results are 
within the uncertainties due to count fluctuations for most 
of the samples.

Fig. 9 shows the counts  derived by Gardner et al. (1996, 
Gar97), McLeod et al. (1995, McL95), Glazebrook et al. (1994, Gla94), 
Djorgovski et al. (1995, Djo95), Soifer et al. (1994, Soi94), Moustakas
et al. (1997, Mou97) Gardner et al. (1993, HWS, HMWS, HMDS, HDS) 
and Mobasher et al. (1986 Mob86).
The error bars represent  $\sqrt{N}$ statistics.
We shall distinguish three different domains: 
$K<16$, $16<K<20$ and $K>20$.

At bright magnitudes  ($K<16$),  the counts follow a linear
$dlogN/dm$ relation with a slope very close to the Euclidean one 
($dlogN/dm=$0.6).
In this magnitude range the counts of  Gardner et al. (1996),
Glazebrook et al. (1994)  and Mobasher et al. (1986) agree very well,
 while a larger scatter is introduced by the HWS and HMWS ones (Gardner et al. 
1993; Gardner 1995a).

In the intermediate magnitude range ($16<K<20$)  a bump with respect 
to both  the brighter and the fainter magnitude ranges is visible.
Such a feature is not present  in the $B$, $R$ and $I$-band galaxy counts
at comparable magnitudes (e.g. Tyson 1988; Metcalfe et al. 1991; 
Metcalfe et al. 1995, Driver et al. 1995).
Gardner et al. (1993) interpret this feature as a flattening
of the count slope to 0.26 at $K>17$, which implies a remarkable 
blueing  trend of  galaxies at these   magnitudes (Cowie et al. 1995).
On the other hand, in the same magnitude range, the galaxy counts 
of McLeod et al. (1995)  suggest a steeper slope (0.32) and
the deep counts of Soifer et al. (1994),  
although systematically higher than the others,  confirm such steeper
 slope.
  
At faint magnitudes ($K>20$) the $dlogN/dm$ is not very well defined,
as shown by the quite different shape and surface density 
derived from the HDS and the Djorgovski et al. (1995) data:
while the HDS counts linearly rise up to $K\sim19.5$ 
and then clearly  decline, 
the Djorgovski et al. counts  continue to rise with a power-law slope of 0.32 
with no evidence of a turnover or of a 
flattening down to the limits of the survey ($K\sim24$).

%\subsection{Comparison with other Counts}
Fig. 10 shows ESOKS1 and
  ESOKS2 galaxy counts superimposed to  those in the literature.
At magnitudes brighter than $K'\sim$16.5 our counts perfectly agree
 with those of Glazebrook et al. (1994), Gardner et al. (1996), 
Mobasher et al. (1986) and the HMWS; note that in this magnitude range
 our surveys cover a 
much smaller area than the other authors'.
At fainter magnitudes, where we surveyed a larger area than other
authors did,  our counts are systematically lower than 
those of HMDS and HDS,   McLeod et al. (1995) and  Soifer et
 al. (1994) and  do not show the bump found by  these surveys at
these magnitudes.   

Many factors may contribute to justify differences between counts derived
from different surveys: field selection criteria, 
magnitude estimate,  correction for star contamination,
 width and sampling of the surveyed areas.
The bias produced by a blank field selection criterion,
often adopted  in faint galaxy count surveys, significantly 
affects counts only at $K<16$, and it is negligible at fainter
magnitudes (\S 2.1).
In principle, different methods of estimating magnitudes could make
a comparison more difficult.
On the other hand, magnitudes can all be reconducted to a ``total''
magnitude (e.g. by using standard growth curves), even if
some uncertainties are necessarily introduced.
On the contrary, the correction for star contamination 
(e.g., through color-color diagrams)  depends on the filters used  and
different results could be  affected by different effects.
And, last but not least, different  area widths are affected by count
fluctuations and, for a given surveyed area,  fluctuations may 
depend on the number of fields which sample the area and on their
  distribution on the sky (sparse or adjacent fields).

The number counts of Soifer et al. (1994),
obtained over an area of $\sim2.4$ arcmin$^2$, appear to be significantly
higher than  the others, in particular the difference between our and
their surface density of galaxies  in the magnitude range  $16<K'<19$ 
is $\sim12000$ galaxies per square degree ($\sim8$ galaxies on 2.4
arcmin$^2$). 
Soifer et al.'s counts were obtained surveying areas surrounding 
high-redshift target
 objects and they are not corrected for star contamination.
Magnitudes are estimated within $3''$ and corrected to total magnitude
by  summing the mean value of the difference between the $3''$ and 
a larger aperture magnitude for a sample of  bright objects.
We  tried to reproduce this method by  estimating  
the $K'_{3''_{corr}}$ magnitude for the sources of the ESOKS2, 
summing the median value of the difference $K'_{5''}-K'_{3''}$ 
to the 3'' aperture magnitude.
In Fig. 11  the number counts obtained using the $K'_{3''_{corr}}$
magnitude are compared to those obtained with FOCAS $K'_{tot}$
magnitude.
The agreement between the two distributions is quite good, thus
showing that the large discrepancy between our counts and those 
derived by  Soifer et al. (1994) cannot be ascribed to different
magnitude estimates.
In an attempt to establish if the observed count fluctuations
can be responsible of such excess obtained over an area of  
$2.4$ arcmin$^2$, from the 38 ESOKS2 high galactic latitude fields
we  randomly extracted  a number of fields
corresponding to such area (one field in this case).
By repeating N times the procedure and computing the surface
density of galaxies each time, we  estimate that 
 the probability to obtain the same number counts of Soifer et al. (1994)
 over an area of $2.4$ arcmin$^2$  is 4\%.
This low probability is marginally significant, 
but taking into account  the star contamination which affects the 
Soifer et al.'s sample,  the surface density
is probably compatible with count fluctuations.
We feel however that, in this magnitude range, such high surface density  may reflect a real high density region,
maybe related to the peculiar selection criterion adopted
for the target areas and the missing correction for star contamination.

McLeod et al. (1995) used FOCAS total magnitudes, as
we did. 
They surveyed an area of $\sim22$ arcmin$^2$ divided in two separate fields.
Star-galaxy separation has been done on the basis of 
 a color-color diagram.
They count an excess of 5600 galaxies per square degree (34 galaxies
on 22 arcmin$^2$) with respect
to our data in the  magnitude range $16<K'<19$.
We estimated that the probability to observe the surface density found
by McLeod et al. (1995) over $\sim22$ arcmin$^2$,
corresponding to $\sim7$ ESOKS fields, is  3\%, which
implies that the observed count fluctuations alone cannot justify the
 discrepancy between our and their counts.

In the case of HMDS (Gardner 1995a), which shows the highest
surface density of galaxies down to $K\sim18$,
the excess with respect to our data is $\sim6600$ gal/deg$^2$
in the magnitude range $16<K'<19$ .
Taking into account that we surveyed approximately 
the same area down to $K'=18$ ($\sim170$ vs $\sim160$ arcmin$^2$), 
such discrepancy cannot be  explained in terms of
count fluctuations only.
In the HMDS sample, the magnitudes are calculated 
within an aperture of 6$\arcsec$ which is not significantly 
different from an aperture of $5''$.
In Fig. 2 we did show that the $K'_{5^{\arcsec}}$ and $K'_{tot}$
magnitude distributions do not differ, as confirmed also by
a KS-test, thus implying that the different counts 
are not due to the different magnitudes used.
Thus, in this case the difference cannot be reconduced to any
of the above mentioned factors.
 
The discrepancy between the ESOKS2 and the HDS is smaller than 
with other surveys, the deviations being of the order of $1\sigma$ in
each magnitude bin.
 The discrepancy is larger when compared to
the ESOKS1 sample.
The area covered by the HDS is $\sim16.5$ arcmin$^2$ and 
their star-galaxy separation  is based on the
inverse second moment of the light distribution (Kron 1980)
and the color-color diagram.
Magnitudes are estimated within $3''$ and corrected to total magnitude
by summing the median value of the difference computed 
between $6''$ and $3''$ aperture magnitudes for a sample of isolated
objects (Cowie et al. 1994).
We have already shown  how the different magnitudes 
used cannot explain the observed discrepancies (see Fig. 10).
The HDS surface density of galaxies in the magnitude range
$16<K<19$ is a factor $\sim1.35$ larger than that derived from the ESOKS2 
(corresponding to an excess of 3900 galaxies per square degree) and  a 
factor 1.7 larger than  that obtained from the ESOKS1.
In terms of pure count fluctuations,
the probability to observe such surface density 
 over $\sim16.5$ arcmin$^2$ is 10\%.
Thus  the discrepancy between our counts and those derived by the HDS
is consistent with  count fluctuations.

\section{Summary and Conclusions}
We have presented galaxy counts in the magnitude range $13<K'<20$, 
derived from two  independent surveys 
covering $\sim40$ arcmin$^2$ (ESOKS1) and $\sim170$ arcmin$^2$ 
(ESOKS2) in the Southern Hemisphere.
Our counts are in quite good agreement with those of 
other authors  at bright magnitudes ($K'<16-17$),
and  follow a $dlogN/dm$ relation with a slope of 0.57.
At fainter magnitudes ($K'>17$), our counts 
are systematically lower than  the others and do not show 
the bump seen in other $K$-band surveys.
Such a discrepancy cannot  be accounted for  in terms  of the 
different magnitudes used while could be consistent with count fluctuations in most cases. 
On the other hand  the systematic nature of the discrepancy
makes significant the absence of the bump in our data at
magnitudes $K'>17$.
It is more difficult to quantify the uncertainty in our counts
due to the methods we were forced to use in order to subtract the
star contribution, even if at $K'>17$  
the expected number of stars is relatively low. A clue is given
by the source counts
(i.e. galaxies and stars) which we obtained on the 16 high galactic latitude fields (see Table 5): these counts are necessarily
upper limits to the galaxy counts and
they unambiguously confirm the discrepancy with other authors and
exclude its dependency from the uncertainties in our star correction
methods.   
 
Our counts support the conclusions of Gardner et al. (1993) of a
changing in the slope of the $dlogN/dm$ relation at $K'\sim17$,
confirming the blueing trend of faint galaxies shown by
 Cowie et al. (1995).
However, we measure a $dlogN/dm$ slope  (0.35 at $K'\ge17$) 
 in agreement with those found by McLeod et al. (1995)
and  Djorgovski et al. (1995)
but higher than the value of 0.26 found by Gardner et al. (1993).

In the  ESOKS2 subsample  limited to $K'\sim18$, we detect
a significant fraction (10\%) of very blue galaxies ($B-K'<3$).
This population  is not present in  other $K$-band
selected samples.
Such deficiency is expected in surveys where a 
   blank-field selection
criterion has been applied (e.g. HMDS; HDS; ESOKS1; McLeod et al. 1995).
The very blue $B-K'$ and $B-R$ colors of such a population indicates 
that they are star-forming galaxies.
From the $B-K$ vs $M_K$ relation (Saracco et al. 1996) 
we infer that 
they  have low luminosities and are at redshifts $z<0.27$.
They may represent a  population of blue sub-$L^*$ galaxies 
similar to that previously suggested by Lilly et al. (1995)
at  $z<0.2$.

\acknowledgements
We are grateful to A. Moorwood and C. Lidman for advice on the 
use of IRAC2 and to L. L. Cowie for suggestions on the data 
analysis procedures.
Special thanks are due to M. Cohen, who provided us with the
output of his SKY galaxy model suited to our fields.

\newpage
%\section{Figure Captions}
\figcaption{Total FOCAS magnitude distribution ($K'_{tot}$) for the 
ESOKS1 sample}
\figcaption{Distribution of total  and aperture magnitudes
 of the ESOKS2 sample.}
\figcaption{$B-R$ vs. $R-K'$ color-color plot for the optically
classified objects (stars represent stars and open circles galaxies)
and for the unclassified objects (crosses).}
\figcaption{$K'$ isophotal detection area vs. $K'_{tot}$ magnitude.
The symbols are as in Fig. 3.}
\figcaption{Complement to 1 of the ratio between the number of objects which FOCAS classifies as stars on the IR frames ($S_K$) and on the $B$ and $R$ frames ($S_{BR}$) as a function
of $K'$ magnitude.}
\figcaption{$K'$-band galaxy number counts derived from the ESOKS1 (starred
symbols) and ESOKS2 (filled circle).}
\figcaption{Color-magnitude diagram for the $K'\le18$ galaxies
(open circles), the $K'\le18$ unclassified sources (crosses),
and the $K'>18$ unclassified objects (filled circles).
Given that the B limiting  magnitude differs from field to field,
the two continuous lines represent the minimum and the maximum
value of $B_{lim}$.}
\figcaption{$B-K$ color distribution of the $K\le18$ ESOKS2 sample
(solid thick line), the HMDS-SSA (dashed line),  and the HMDS-Dur (dotted line) samples. The distributions are normalized to
the same number of sources. The dashed histogram includes
the lower limits to the $B-K$ color of the 10 ESOKS2 
optically undetected sources.}
\figcaption{$K$-band galaxy number counts from the literature.
The continuous line shows a slope $dlogN/dm=0.6$.}
\figcaption{ESOKS1 and ESOKS2 galaxy number counts superimposed to those
from the literature.}
\figcaption{Galaxy number counts obtained by using total FOCAS magnitude
are compared to those obtained by using $K'_{3''_{corr}}$ magnitudes.}


\begin{thebibliography}{}
\bibitem{}{}Abraham R. G., Tanvir N. R., Santiago B. X., Ellis R. S., Glazebrook K. \& van den Bergh S. 1996, \mnras,    279, L47
\bibitem{}{} Avni Y., Soltan A., Tanambaum H. \& Zamorani G. 1980, \apj,235, 694
\bibitem{} Banchet V., Mamon G. A. \& Contensou M 1997, Astr. Lett. \& Comm., in press
\bibitem{} Bernstein G. M., Nichol R. G., Tyson J. A., Ulmer M. P. \&
Wittman D. 1995, \aj,   110, 1507
\bibitem{} Bershady M. A. et al. 1994, \aj,   108, 870
\bibitem{}{} Bertin E. \& Dennefeld M. 1997, \aap,    317, 42
\bibitem{} Brainerd T. G., Smail I. R. \& Mould J. R. 1994, BAAS 185, 114.06
\bibitem{} Broadhurst T. J., Ellis R. S. and Glazebrook K. 1992, {Nat. 355, 55}
\bibitem{} Cohen M. 1993, AJ 105, 1860
\bibitem{} Cohen M. \& Saracco P. 1997, in preparation
\bibitem{}{} Colless M. M., Ellis R. S., Broadhurst T. J., Taylor K. \& Peterson B. A. 1993, \mnras,    261, 19
\bibitem{} Colless M. M., Schade D., Broadhurst T. J. \& Ellis R. S. 1994, \mnras,    267, 1108
\bibitem{}{} Cowie L. L., Gardner J. P., Lilly S. J. \& McLean I. 1990, \apj,  360, L1
\bibitem {}Cowie L. L., Songaila A. and Hu E. M. 1991, {Nat. 354, 460}
\bibitem{}{} Cowie L. L., Gardner J. P., Hu E. M., Songaila A., Hodapp K. W. \& Wainscoat R. J. 1994, \apj, 434, 114
\bibitem{}{} Cowie L. L., Hu E. M. \& Songaila A. 1995, \aj,   110, 1576
\bibitem{}{} Djorgovski et al. 1995, \apj, 438, L13
 \bibitem{}{} Driver S. P., Windhorst R. A. \& Griffiths R. E. 1995, \apj, 453, 48
\bibitem{}{} Driver  S. P., Couch W. J., Phillipps S. \& Windhorst R. A. 1996, \apj, 466, L5
\bibitem{}{} Elias J. H., Frogel J. A., Matthews K. \& Neugebauer G. 1982, \aj,   87, 1029
\bibitem{}{} Feigelson E. D. \& Nelson P. I. 1985, \apj, 293, 192
\bibitem{}{} Gardner J. P., Cowie L. L. \& Wainscoat R. J. 1993, \apj, 415, L9
\bibitem{}{} Gardner J. P. 1995a, \apjs 98, 441
\bibitem{} Gardner J. P. 1995b, \aj,   452, 538
\bibitem{}{} Gardner J. P., Sharples R. M., Carrasco B. E. \& Frenk C. S.
1996, \mnras, 282, L1   
\bibitem{} Glazebrook K., Peacock J., Collins C. \& Miller L. 1994, \mnras,    266, 65
\bibitem{}{} Glazebrook K., Ellis R., Santiago B. \& Griffiths R. 1995, \mnras,    275, L19
\bibitem{}{} Griffiths R. E. et al. 1994a, \apj, 437, 67
\bibitem{}{} Griffiths R. E. et al. 1994b, \apj, 435, L19
\bibitem{}{} Guiderdoni B. 1991, in Annales de Physique, 16, 235
\bibitem{} Hodapp K. W., Rayner J. \& Irwin E. 1992, PASP 104, 441
\bibitem{} Huang J. S., Cowie L. L., Gardner J. P., Hu E. M., Songaila A., \& Wainscoat R. J. 1997, \apj, 476, 12
\bibitem{}{} Isobe T., Feigelson E. D. \& Nelson P. I. 1986, \apj, 306, 490
\bibitem{} Jarvis J. F. \& Tyson J. A. 1981, \aj,   86, 476
\bibitem{} Jones L. R., Fong R., Shanks T., Ellis R. S. \& Peterson B. A.
1991, \mnras,    249, 481
\bibitem{} Kron R. G. 1980, \apjs 43, 305
\bibitem{} Legget S. K. \& Hawkins M. R. S. 1988, \mnras,    234, 1065
\bibitem{} Lidman C., Gredel R. \& Moneti A. 1995, ESO Report n.2, Aug. 1995
\bibitem{} Lilly S. J., Tresse L., Hammer F., Crampton D. \& Le Fevre O. 1995, \apj, 455, 108
\bibitem{} Maddox S. J., Sutherland W. J., Efstathiou G., Loveday J.
\& Peterson B. A. 1990,  \mnras,    247, 1  
\bibitem{} Mamon G. A. 1995, in Wide-Field Spectroscopy and the Distant Universe,
ed. S. J. Maddox \& Aragon-Salamanca (Singapore: World Scientific)
\bibitem{} McLeod B. A., Bernstein G. M., Reike M. J., Tollestrup E. V. \& Fazio G. G. 1995, \apjs 96, 117
\bibitem{} Metcalfe N., Shanks T., Fong R. \& Jones L. R. 1991, \mnras,    249, 498
\bibitem{} Metcalfe N., Shanks T., Fong R. \& Roche N. 1995, \mnras,    273, 257
\bibitem{} Mobasher B., Ellis R. \& Sharpless R. 1986, \mnras,    223, 11
\bibitem{} Morwood A. et al. 1992, Messenger 69, 61
\bibitem{} Moustakas L. A., Davis M., Graham J. R., Silk J.,
Peterson B. A. \& Yoshii Y. 1997, \apj,  475, 44
\bibitem{} Saracco P., Chincarini G. \& Iovino A. 1996, \mnras, 283, 865
\bibitem{} Soifer B. T. et al. 1994, \apj,  420, L1
\bibitem{}  Tyson J. A. 1988,  A.J. 96, 1  
\bibitem{} Tyson J. A. 1989, in CCDs in Astronomy, ed. G. H. Jacoby, A.S.P. Conference Series 8, 1
\bibitem{} Valdes F. 1982, in Faint Object Classification and Analysis System
(Tucson: NOAO)
\bibitem{} Wainscoat R. J. \& Cowie L. L. 1991, \aj,   103, 332
\bibitem{} White S. D. M. 1989, in The Epoch of Galaxy Formation, ed. C. S. Frenk, R. S. Ellis, T. Shanks, A. F. Heavens \& J. A. Peacock (Dordrecht: Kluwer), 15
\end{thebibliography}
\end{document}